\documentclass[preprint,tightenlines,aps,floats,nofootinbib,showpacs]{revtex4}

\usepackage{epsf}
\usepackage{epsfig}
\usepackage{graphicx}
\usepackage[dvips]{color}

\newcommand{\notE}{ \hbox{{$E$}\kern-.60em\hbox{/}}}
\newcommand{\notp}{\ \hbox{{$p$}\kern-.43em\hbox{/}}}
\def\D0{\mbox{D\O}}

\newcommand{\eps}{\epsilon}

\includegraphics{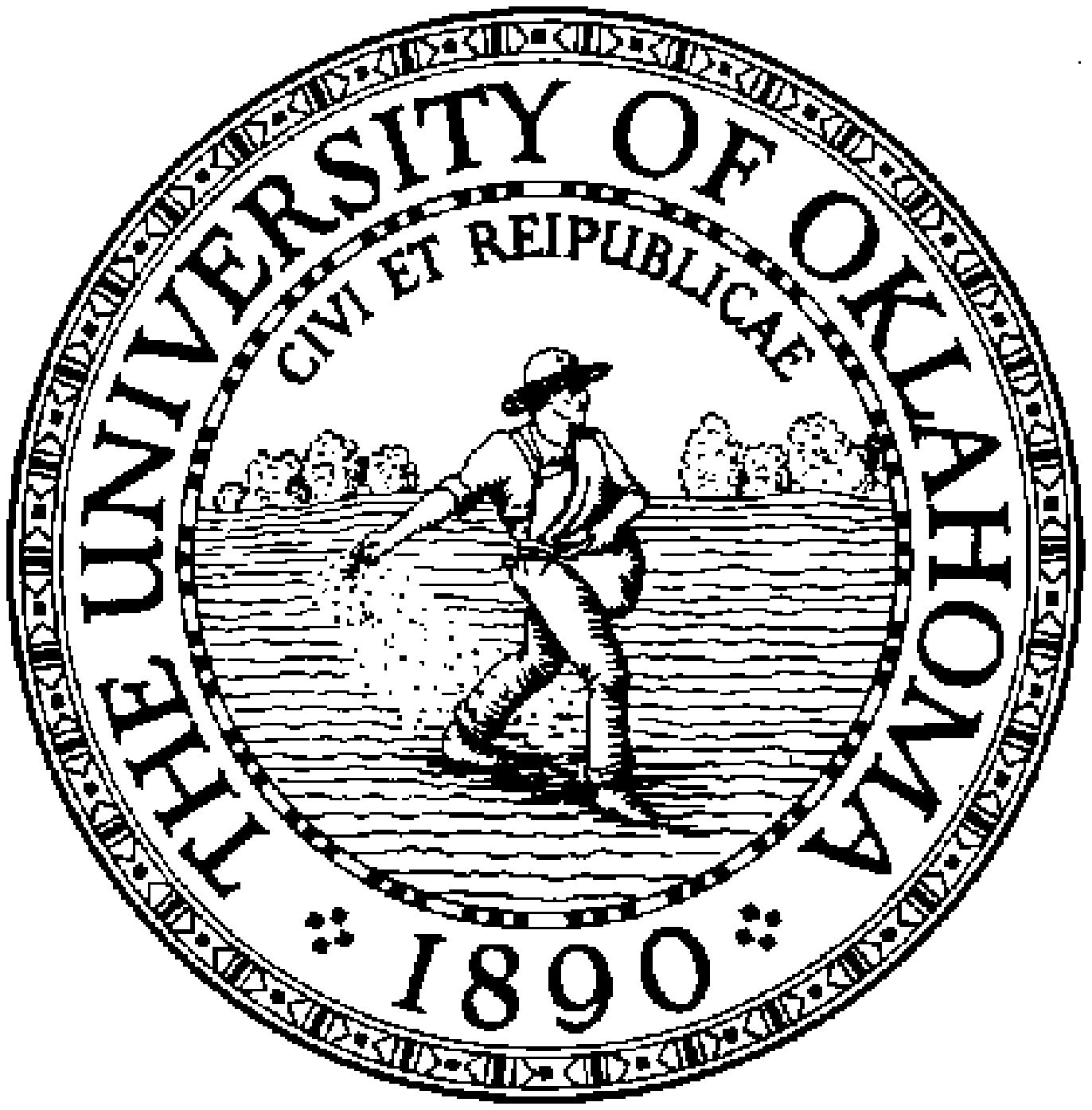}

\preprint{\font\fortssbx=cmssbx10 scaled \magstep2
\hbox to \hsize{
\hskip1.2in %\raise.1in
\hbox{\fortssbx The University of Oklahoma}
\hskip0.2in $\vcenter{
                      \hbox{\bf arXiv: [hep-ph]}
                      \hbox{\bf OU-HEP-110923}
                      \hbox{August 2012}}$ }
}

\begin{document}

%\draft

%-----------------------------------
% Title
%-----------------------------------
\title{\vspace*{0.7in}
Top Decays with Flavor Changing Neutral Higgs Interactions \\ at the LHC}

%-----------------------------------
%   Authors
%-----------------------------------
\author{
Chung Kao$^{a,b}$\footnote{E-mail address: kao@physics.ou.edu},
Hai-Yang Cheng$^b$\footnote{E-mail address: phcheng@phys.sinica.edu.tw},
Wei-Shu Hou$^c$\footnote{E-mail address: wshou@phys.ntu.edu.tw}, and
Joshua Sayre$^a$\footnote{E-mail address: sayre@physics.ou.edu}}

%-----------------------------------
%   Address
%-----------------------------------
\affiliation{
$^a$Homer L. Dodge Department of Physics and Astronomy,
University of Oklahoma, Norman, OK 73019, USA \\
$^b$Institute of Physics, Academia Sinica,
Taipei 11529, Taiwan, ROC \\
$^c$Department of Physics, National Taiwan University,
Taipei 10617, Taiwan, ROC}

\date{\today}

\bigskip

%-----------------------------------
%   Abstract
%-----------------------------------
\begin{abstract}

We investigate the prospects for discovering a top quark decaying into
one light Higgs boson along with a charm quark in top quark pair production
at the CERN Large Hadron Collider (LHC).
A general two Higgs doublet model is adopted to study the signature
of flavor changing neutral Higgs decay $t \to c\phi^0$, 
%or $\bar{t} \to \bar{c}\phi^0$ 
where $\phi^0$ could be CP-even ($H^0$) or CP-odd ($A^0$).
The dominant physics background is evaluated with realistic
acceptance cuts as well as tagging and mistagging efficiencies.
For a reasonably large top-charm-Higgs coupling 
($\lambda_{tc}/\lambda_{t} \agt 0.09$),
the abundance of signal events and the %that our acceptance cuts
reduction in physics background allow us to establish a $5\sigma$ signal for
$M_\phi \sim 125$~GeV 
at the LHC with a center of mass energy ($\sqrt{s}$) of 8 TeV 
and an integrated luminosity of $20$~fb$^{-1}$.
The discovery potential will be greatly enhanced
with the full energy of $\sqrt{s} = 14$ TeV.

\end{abstract}

%------------------------
% PACS numbers
%------------------------
%
%\begin{itemize}
%\item 12.15.Mm: Neutral currents
%\item 12.60.Fr: Extensions of electroweak Higgs sector
%\item 14.65.Ha: Top quarks
%\item 14.80.Ec: Other neutral Higgs bosons
%\end{itemize}

%-----------------------------------------------------------------------
% Make Title Page
%-----------------------------------------------------------------------

\maketitle

%=======================================================================
%   BEGIN MAIN TEXT
%=======================================================================
\newpage

%-----------------------------------------------------------------------
% I. Introduction
%-----------------------------------------------------------------------
\section{Introduction}

The Standard Model has been very successful in explaining
almost all experimental data to date,
culminating in the discovery of the top quark \cite{Abe:1995hr,Abachi:1995iq}
and the tau neutrino \cite{Kodama:2000mp}, and finally, a scalar particle that 
appears to be the long awaited Higgs boson has emerged
at the Large Hadron Collider~\cite{
CMS_Seminar,ATLAS_Seminar,ATLAS_Higgs12,CMS_Higgs12,Higgs_ATLAS7,Higgs_CMS7}.
The most important experimental goals of the CERN Large Hadron Collider (LHC)
are the investigation of the mechanism for electroweak symmetry
breaking (EWSB) --- the discovery of the Higgs bosons or the proof of their
non-existence --- and the search for new physics beyond the Standard Model (SM).
It is important now to check whether the emerging Higgs boson
fits the SM prescription. 

In the Standard Model there is just one Higgs doublet, which generates
masses for vector bosons and fermions, but the differences among Yukawa
couplings of fermions with the Higgs boson are not explained.
In addition, there are no flavor changing neutral currents (FCNC) mediated by
gauge interactions or by Higgs interactions at the tree level.
At present, the top quark is the most massive elementary particle ever
discovered. It might provide clues to study the mechanisms of EWSB and FCNC.
At the one loop level, the branching fraction of $t \to c H$ is
$4.6 \times 10^{-14}$ for $M_H = 120$ GeV~\cite{Mele:1998ag,Eilam:1990zc}.
If this decay mode is detected in the near future, it would likely indicate 
a large effective coupling of tree-level origins, 
or very large enhancement from beyond SM loop effects.

A general two Higgs doublet model usually contains flavor changing
neutral Higgs (FCNH) vertices if there is no discrete symmetry
to turn off tree-level FCNC~\cite{Glashow:1976nt,Guide}.
In the context of invoking~\cite{Cheng:1987rs} a 
Fritzsch-like quark mixing Ansatz~\cite{Fritzsch:1977vd}
to evade low energy FCNC constraints induced by FCNH couplings,
it was pointed out long ago~\cite{Hou:1991un} that 
top-charm FCNH coupling could be prominent because of the large top mass. 
With this in focus, a special two Higgs doublet model for the top quark (T2HDM)
\cite{Das:1995df} offers a good explanation of why the top quark is
much heavier than other elementary particles.
In the T2HDM, the top quark is the only elementary fermion acquiring
its mass from a special Higgs doublet ($\phi_2$) with
a large vacuum expectation value (VEV).
Since the up and charm quarks couple to another Higgs doublet ($\phi_1$),
there are FCNH interactions among the up-type quarks.

In the near future, for an integrated luminosity of ${\cal L} = 10$ fb$^{-1}$ 
at $\sqrt{s} = 8$ TeV,\footnote{
 The current expectation is that $\sim 20$ fb$^{-1}$ data 
 would be collected by both ATLAS and CMS in 2012~\cite{LHC_Lumi}.
 }
the LHC will produce approximately $2 \times 10^6$ top quark pairs
($t\bar{t}$)~\cite{arXiv:1009.4935,arXiv:1105.5824,arXiv:1111.5869}
for $m_t \simeq 173$ GeV.
For the same integrated luminosity at $\sqrt{s} = 14$ TeV,
the number of ($t\bar{t}$) pairs generated would increase to about $1 \times 10^7$.
Thus, the LHC is becoming a top quark factory, providing great
opportunities to study electroweak symmetry breaking as well as
other important properties of the top quark.
If the top quark is heavier than a neutral Higgs boson that interacts
with a top quark and a charm quark, then a promising FCNH signature
will appear in pp collisions as
$pp \to t\bar{t} \to t\bar{c}\phi^0 +X$ or
$pp \to t\bar{t} \to c\phi^0\bar{t} +X$
at the LHC~\cite{Aguilar-Saavedra:2000aj}.
With the emerging scalar object at 125 GeV, the pursuit of this
signature has become mandatory.

In this letter, we study the discovery potential of the LHC in the search
for the rare top decay $t \to c\phi^0$, where $\phi^0$ is a scalar
($H^0$) or a pseudoscalar ($A^0$). The Higgs boson then decays
into a pair of bottom quarks ($b\bar{b}$).
We have evaluated production rates with full tree-level matrix elements
including Breit-Wigner resonances for both the signal and the physics
background. In addition, we optimize the acceptance cuts to effectively
reduce the background with realistic $b$-tagging and mistagging
efficiencies. Promising results are presented for the LHC
with $\sqrt{s} = 8$ TeV as well as $\sqrt{s} = 14$ TeV.
Section II shows the production cross sections for the Higgs signal
and the dominant background, as well as our strategy to determine
the reconstructed masses for the top quark and the Higgs boson.
Realistic acceptance cuts are discussed in Section III.
Section IV presents the discovery potential at the LHC
for $\sqrt{s} = 8$ TeV and $\sqrt{s} = 14$ TeV.
Our conclusions, which are quite optimistic, are drawn in Section V.

%------------------------------------------------------------------------------
% II. The Higgs signal and the Physics Background
%------------------------------------------------------------------------------
\section{The Higgs Signal and the Physics Background}

\subsection{The Higgs Signal}

We adopt a general two Higgs doublet model to study flavor changing
neutral Higgs interactions with the following effective Lagrangian
\begin{eqnarray}
{\cal L}
 = -\lambda_{tc} \bar{c}tH^0
   -i\lambda_{tc} \bar{c}\gamma_5 t A^0 +{\rm H.c.},
\end{eqnarray}
where $H^0$ is a CP-even scalar, $A^0$ is a CP-odd pseudoscalar and
$v = 2M_W/g_W \simeq$ 246 GeV.
For $M_\phi < m_t$, the $t \to c\phi^0$ decay width \cite{Hou:1991un} is
\begin{eqnarray}
\Gamma(t \to c\phi^0)
 =  \frac{|\lambda_{tc}|^2}{16\pi}\times (m_t)\times
      [ (1\pm\rho_c)^2 -\rho_\phi^2 ]
      \times \sqrt{1-(\rho_\phi+\rho_c)^2}\sqrt{1-(\rho_\phi-\rho_c)^2},
\end{eqnarray}
where $\phi^0 = H^0$ or $A^0$, $\rho_c = m_c/m_t$, $\rho_\phi = M_\phi/m_t$,
and + or - corresponds to $\phi^0$ being a scalar or a pseudoscalar.
We assume that the total decay with of the top quark is
\begin{eqnarray}
\Gamma_t = \Gamma( t\to bW ) +\Gamma( t \to c\phi^0 ) \, .
\end{eqnarray}
Then the branching fraction of $t \to c\phi^0$ becomes
\begin{eqnarray}
{\cal B}(t \to c\phi^0) = \frac{ \Gamma(t\to c\phi^0) }{ \Gamma_t } \, .
\end{eqnarray}

As a case study, we take the FCNH Yukawa couplings
to be the geometric mean of the Yukawa couplings
of the quarks\footnote{
We note that some physicists choose the Yukawa coupling to be
$y_{tc} = \sqrt{2}\lambda_{tc}$.}~\cite{Fritzsch:1977vd,Cheng:1987rs}
\begin{equation}
\lambda_{tc} = \frac{ \sqrt{m_t m_c} }{v} \simeq 0.063,
\end{equation}
with $m_t = 173.3$ GeV and $m_c = 1.4$ GeV.
Then the branching fraction of $t \to c\phi^0$ becomes
${\cal B}(t\to c\phi^0) = 2.2 \times 10^{-3}$ for $M_\phi = 125$ GeV or
${\cal B}(t\to c\phi^0) = 6.2 \times 10^{-4}$ for $M_\phi = 150$ GeV. For
illustration we use this ansatz in all our figures except for Fig.~5.
Later, in the section on LHC discovery potential, we will consider
$\lambda_{tc}$ as a free parameter. We assume that the width and branching
fraction of the Higgs scalar decay to $b\overline{b}$ are similar to
those of the standard Higgs boson, while the Higgs pseudoscalar
decays to the $W^+W^-$ or the $ZZ$ pairs are negligible. We do not consider CP
violation in this study.

We employ the programs MadGraph~\cite{Stelzer:1994ta,Alwall:2007st} and
HELAS~\cite{Murayama:1992gi} to evaluate the exact matrix element
for the FCNH signal in top decays from gluon fusion
and quark-antiquark annihilation,
$gg \to t\bar{t} \to t\bar{c}\phi^0 \to b\ell^+\nu \bar{c} b\bar{b}$ and
$q\bar{q} \to t\bar{t} \to t\bar{c}\phi^0 \to b\ell^+\nu \bar{c}b\bar{b}$
as well as
$t\bar{t} \to c\phi^0\bar{t} \to cb\bar{b} \bar{b}\ell^-\bar{\nu}$,
where $\ell = e$ or $\mu$.
The signal cross section at the LHC for
$pp \to t\bar{t} \to tc\phi^0 \to b\ell\nu cb\bar{b} +X$
is evaluated with the parton distribution functions of CTEQ6L1~\cite{CTEQ6}.
In addition, we have checked the signal cross section by narrow width
approximation. That is, the cross section
$\sigma(pp \to t\bar{t} \to tc\phi^0 \to b\ell\nu cb\bar{b} +X)$
is calculated as the product of cross section times branching
fractions:
$\sigma(pp \to t\bar{t} \to b\ell\nu \bar{t} +X)
\times {\cal B}(t\to c\phi^0) \times {\cal B}(\phi^0\to b\bar{b})$.
The factorization scale and the renormalization scale are chosen to be
$Q = M_{t\bar{t}}$, the invariant mass of $t\bar{t}$.
This choice of scale leads to a K factor of 2 for top quark pair
production~\cite{Bonciani:1998vc,Nason:1987xz}.

In our analysis, we consider the FCNH signal from both
$t\bar{t} \to t\bar{c}\phi^0 \to b\ell^+\nu \bar{c} b\bar{b}$
and
$t\bar{t} \to c\phi^0\bar{t} \to cb\bar{b} \bar{b}\ell^-\bar{\nu}$,
which will be commonly described as
$t\bar{t} \to c\phi^0\bar{t} \to b\ell\nu cb\bar{b}$ or 
$t\bar{t} \to c\phi^0\bar{t} \to b\ell\nu b\bar{b}j$.
In every event, we require there should be 3 $b$ jets and one light jet
($j = u, d, s, c$, or $g$).
In addition, there is an isolated lepton ($\ell = e$ or $\mu$),
and the neutrino will lead to missing transverse energy ($\notE_T$).
Unless explicitly specified,
$q$ generally denotes a quark ($q$) or an anti-quark ($\bar{q}$)
and $\ell$ will represent a lepton ($\ell^-$) or anti-lepton ($\ell^+$).
That means our FCNH signal leads to the final state of
$b\ell\nu b\bar{b}j$ or $bbbj\ell +\notE_T$.

%------------------------
% Physics Background
%------------------------
\subsection{The Physics Background}

The dominant physics background to the final state of $bb\bar{b}j \ell\nu$
comes from top quark pair production followed by top and $W$ decays:
$pp \to t\bar{t} \to b\ell^+\nu \bar{b}\bar{c}s +X$ or
$pp \to t\bar{t} \to bc\bar{s} \bar{b}\ell^-\bar{\nu} +X$,
where a $c$-jet is mis-identified as a $b$-jet.
We have also considered backgrounds from
$t\bar{t} \to b\ell\nu \bar{b} u\bar{d}$,
as well as backgrounds from the production of
$b\bar{b} b\bar{b} \ell\nu$ and $b\bar{b} c\bar{c} \ell\nu$.
According to the ATLAS and CMS Technical Design Reports \cite{ATLAS,CMS},
the $b$ tagging efficiency is $50\% - 60\%$,
the probability that a $c$-jet is mistagged as a $b$-jet ($\epsilon_c$)
is approximately $10\%$, while
the probability that any other jet is mistagged as a $b$-jet ($\epsilon_j$)
is $1\%$.
If $\eps_c$ is chosen to be the same as that of a light jet ($\eps_j = 0.01$),
the background from $t\bar{t} \to b\ell\nu b c s$ will be
underestimated by a factor of 10 such as
in the analysis in Ref.~\cite{Aguilar-Saavedra:2000aj}.
This point would be elaborated in our study.

%------------------------
% Mass Reconstruction
%------------------------
\subsection{Mass Reconstruction}

In this subsection, we discuss our strategy to determine
the reconstructed top mass as the invariant mass of $M_{bbj}$
for the top quark with FCNC
($t \; {\rm or} \; \bar{t} \to c\phi^0 \to cb\bar{b}$),
as well as the other reconstructed top mass with leptonic decay
($\bar{t} \; {\rm or} \; t \to bW \to b\ell\nu$):
\begin{eqnarray}
M_{t_1}^R & = & M_{bbj} \, , \\
M_{t_2}^R & = & M_{b\ell\nu} \, .
\end{eqnarray}
In the process of doing so, we reconstruct the Higgs mass as the
invariant mass of a pair of $b$ jets and one $W$ mass as the invariant
mass of a charged -lepton/neutrino pair $\ell \nu$. We also reconstruct a
potential second $W$ mass from a $b$-jet/light jet pair for vetoing the
background:
\begin{eqnarray}
M_\phi^R & = & M_{bb} \, , \\
M_{W_1}^R & = & M_{bj} \, , \\
M_{W_2}^R & = & M_{\ell\nu} \, .
\end{eqnarray}

Figure~1 shows invariant mass distributions with basic cuts:
$p_T(b,j) \ge 15$ GeV, $p_T(\ell) \ge 20$ GeV, 
$|\eta(b,j,\ell)| \le 2.5$ and $\notE_T > 20$ GeV.
In each event, we assume that three $b$ jets and one non-$b$ jet are
identified through $b$-tagging.
We then assign the three $b$ jets ($b_1,b_2,b_3$)
according to the following procedure:
Since our FCNC signal comes from
$t\bar{t} \to c\phi^0 \bar{b}\ell\bar{\nu}
\to b\bar{b}c \bar{b}\ell\bar{\nu} \to bbbj\ell+\notE_T$,
we will choose the pair of $b$ jets that minimize $|M_{bbj}-m_t|$
as $b_1 b_2$ and label the other $b$ jet as $b_3$.
For a correctly reconstructed event, $b_1$ and $b_2$ are the products
of a Higgs decay as well, such that their invariant mass
has a peak near $M_\phi$.
For a background event, one $b$ is likely coming from the top decay
$t \to bW \to bcj$ while the other is either a mistagged $c$ or
a light quark jet coming from $W$ decay, or a real $b$ quark coming
from the decay of $\overline{t}$.
Let us identify $b_2$ as the member of this pair
that minimizes $M_{bj}-m_W$.
In a good reconstruction, the remaining $b$ quark, $b_3$ should
reproduce the top quark mass with the charged lepton and neutrino momenta.
In this figure, we present the reconstructed masses for signal and
background with $M_\phi = 125$ GeV: $M_{bbj} = M_{b_1 b_2 j} = M_{t_1}^R$,
$M_{bb} = M_{b_1 b_2} = M_\phi^R$, $M_{bj} = M_{b_2 j} = m_{W_1}^R$,
and $M_{b\ell\nu} = M_{b_3 \ell\nu} = m_{t_2}^R$.
We have used the Higgs scalar case here and in Fig.~2 as an
example; the shapes are virtually the same for a pseudoscalar.

%------------------------------------------------
% FIG. 1
%------------------------------------------------

\begin{figure}[t!]
\centering\leavevmode
\epsfxsize=4.0in
\epsffile{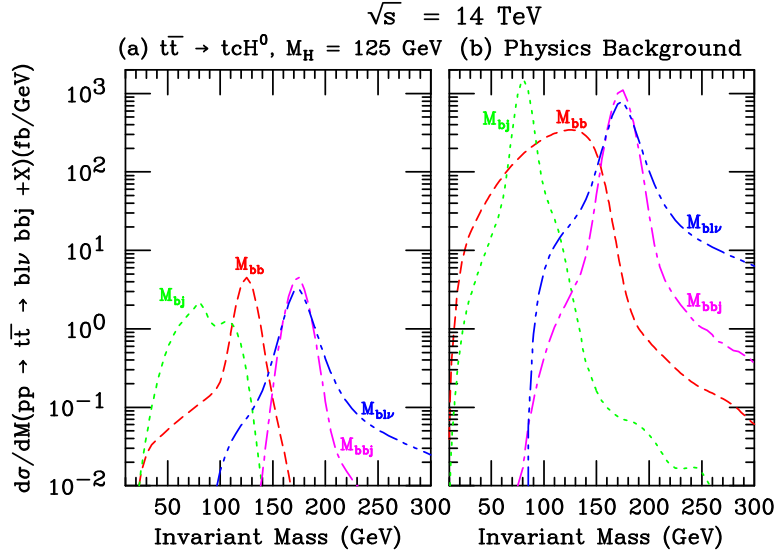}
\caption[]{
Invariant mass distribution of $bbj$ (dash-dot magenta),
$b\ell\nu$ (dash-dot-dot blue), $bb$ (dash red), and $bj$ (dot green),
with basic cuts on $p_T$ and $\eta$ for
(a) the Higgs signal from
$pp \to t\bar{t} \to t cH^0 \to b\ell\nu c b\bar{b} +X$ and
(b) the Standard Model background from
$pp \to t\bar{t} \to b\ell\nu bcs +X$ where $q$ is a $q$ or a $\bar{q}$.
\label{fig:mass} }
\end{figure}

In our analysis, we assume that the FCNH signal comes from top quark
pair production with one top quark decaying via FCNC while the other
decays leptonically ($t \to bW \to b\ell\nu$).
For a real $W$ decay ($W \to \ell\nu$), the momentum of the neutrino ($k$)
and that of the lepton ($p$) have the following relation
\begin{eqnarray}
 (k+p)^2 = m_W^2 \, .
\end{eqnarray}
If the missing transverse energy comes solely from the neutrino in $W$ decay,
we can estimate the longitudinal momentum of the neutrino
($k_z$) with measured lepton energy and momentum $(E_\ell,\vec{p})$,
transverse missing momentum ($k_x = \notE_x$ and $k_y = \notE_y$) and
$E_\nu = |\vec{k}|$.

Assuming an on-shell $W$, we can evaluate $k_z$ of the neutrino with
\begin{eqnarray}
k_z & = & \frac{p_z[2(p_xk_x+p_yk_y) +m_W^2 -m_\ell^2]
           \pm E_\ell \Delta }{m_\ell^2+p_T^2},
\end{eqnarray}
with
\begin{eqnarray}
\Delta^2  =
  [2(p_xk_x+p_yk_y) +m_W^2 -m_\ell^2]^2 -4k_T^2(m_\ell^2+p_T^2) \, .
\end{eqnarray}
There are two possible values for $k_z$ if $\Delta^2 > 0$.
We select whichever leads to a better reconstruction of the top-quark mass,
\begin{eqnarray}
  \text{Min}[m_t^2 -(k+p+p_{b_3})^2],
\end{eqnarray}
and define this reconstructed top mass as
 $M_{t_2}^R = M_{b_3 \ell\nu}$.
In case $\Delta^2 < 0$, we set $\Delta = 0$ to evaluate $k_z$,
corresponding to a virtual $W$.  We require that the reconstructed $W$
mass ($M_{\ell\nu}$) should be close to the on-shell mass $m_W$.

%------------------------------------------------------------------------------
% III. Realistic Acceptance Cuts
%------------------------------------------------------------------------------
\section{Realistic Acceptance Cuts}

To study the discovery potential of this FCNH signal at the LHC,
we have applied realistic cuts and tagging efficiencies for three
combinations of the CM energy ($\sqrt{s}$) and integrated luminosity ($L$):
(a) the early stage of LHC with $\sqrt{s} = 8$ and
$L =$ 5--20 fb$^{-1}$~\cite{LHC_Lumi},
(b) full CM energy ($\sqrt{s} = 14$ TeV) with low luminosity
$L = 30$ fb$^{-1}$, and
(c) full CM energy ($\sqrt{s} = 14$ TeV) with high luminosity
$L = 300$ fb$^{-1}$ \cite{ATLAS,CMS}.

For (a) the early LHC and (b) full CM energy with low luminosity, we require
that in every event there should be
\begin{itemize}
\item
exactly 4 jets that have $p_T > 15$ GeV and $|\eta| < 2.5$,
and three of them must be tagged as $b$-jets;
\item
exactly one isolated lepton that has $p_T > 20$ GeV and $|\eta| < 2.5$;
\item
the missing transverse energy ($\notE_T$) must be greater than 20 GeV;
\item
at least one pair of $b$-jets such that the invariant mass of
$b_1 b_2j$ should be near $m_t$: $|M_{b_1 b_2j} -m_t| \le 25$ GeV;
\item
the pair of $b$-jets, $b_1$$b_2$, that reconstructs the top quark 
with FCNH decay should also satisfy
$|M_{b_1b_2} -M_\phi| \le 0.15 M_\phi$;
\item
a third $b$ jet such that the invariant mass of
$b_3\ell\nu$ should be near $m_t$: \\
$|M_{b_3\ell\nu} -m_t| \le 25$ GeV;
\item the reconstructed leptonic $W$ must satisfy
$|M_{\ell\nu} -m_W| \le 15$ GeV.
\end{itemize}

Additionally, to effectively reduce backgrounds from $W \to jj$, we require
$|M_{b_2j}-m_W| > 15$ GeV. We also require
$\Delta R = \sqrt{\Delta\phi^2  +\Delta\eta^2} > 0.4$
between every pair of jets and between each jet and the charged
lepton, to limit QCD production of multi-jets and ensure good
reconstruction of isolated jets and the charged lepton.

In the early stage of LHC with $\sqrt{s} = 8$ TeV, the $b$-tagging
efficiency ($\epsilon_b$) is taken to be $50\%$, the probability that
a $c$-jet is mistagged as a $b$-jet ($\epsilon_c$) is $10\%$ and
the probability that any other jet is mistagged as a $b$-jet ($\epsilon_j$)
is taken to be $1\%$.
At the full CM energy ($\sqrt{s} = 14$ TeV) with an integrated
luminosity ($L$) of 30 fb$^{-1}$, we follow the tagging and mistagging
efficiencies in the ATLAS Technical Design Report~\cite{ATLAS}:
$\epsilon_b = 60\%$, $\epsilon_c = 14\%$ and $\epsilon_j = 1\%$.

For the full CM energy ($\sqrt{s} = 14$ TeV) with a high integrated
luminosity of 300 fb$^{-1}$,  we require $p_T(b,j) > 30$ GeV,
$p_T(\ell) > 20$ GeV, $|\eta(b,j,\ell)| < 2.5$, and $\notE_T > 40$ GeV.
The tagging and mistagging efficiencies are taken to be
$\epsilon_b = 50\%$, $\epsilon_c = 14\%$ and $\epsilon_j = 1\%$.

Furthermore, a powerful acceptance cut on the charm-jet energy was
proposed in a study to search for FCNH top decays at linear
colliders~\cite{Han:2001ap}.
In the rest frame of the top quark, the energy of the charm jet from
$t \to c\phi^0$ is
\begin{eqnarray}
E_c^R =
 \frac{m_t}{2}\left( 1 +\frac{m_c^2}{m_t^2} -\frac{M_\phi^2}{m_t^2} \right)
   \, .
\end{eqnarray}
For $M_\phi = 125$ GeV, $E_c^R \simeq 42$ GeV, while for $M_\phi =
150$ GeV, $E_c^R \simeq 22$ GeV. In the background, the non-$b$ jet, which is
most likely not a charm quark and which arises from $W$
decay, has a distribution of energy which is more spread
out~\cite{Han:2001ap}.

In Fig.~2, we present the distribution with respect to the charm
energy in the top rest frame ($E_c^R$) for
$pp \to t\bar{t} \to b \ell \nu  c b\bar{b} +X$
with (a) $M_H = 125$ GeV and (b) $M_H = 150$ GeV.
Also shown is the same distribution for the physics background
from $pp \to t\bar{t} \to b \ell \nu bcs +X$ where the charm quark
is mistagged as a $b$ and the strange quark fakes a charm jet.
In our complete analysis, we choose
32 GeV $< E_c^R <$ 52 GeV for $M_\phi = 125$ GeV, and
12 GeV $< E_c^R <$ 32 GeV for $M_\phi = 150$ GeV.

%------------------------------------------------
% FIG. 2
%------------------------------------------------

\begin{figure}[t!]
\centering\leavevmode
\epsfxsize=4.0in
\epsffile{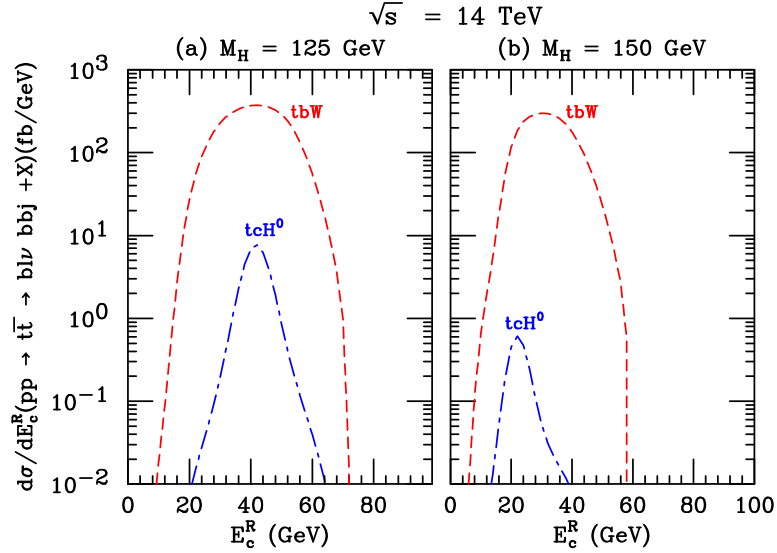}
\caption[]{
Distribution with respect to the charm energy in the top rest frame
($E_c^R$)
$pp \to t\bar{t} \to b \ell \nu  c b\bar{b} +X$ (dot-dash blue)
for (a) $M_H = 125$ GeV
and (b) $M_H = 150$ GeV.
Also shown is the same distribution for the physics background
from $pp \to t\bar{t} \to b \ell \nu bcs +X$ (dash red).
\label{fig:echarm} }
\end{figure}

Measurement uncertainties in jet and lepton momenta as well as missing
transverse momentum give rise to a spread in the reconstructed masses
about the true values of $m_t$ and $M_\phi$.
Based on the ATLAS~\cite{ATLAS} and the CMS~\cite{CMS} specifications
we model these effects by Gaussian smearing of momenta:
\begin{eqnarray}
\frac{\Delta E}{E} = \frac{0.60}{\sqrt{E({\rm GeV})}} \oplus 0.03,
\end{eqnarray}
for jets and
\begin{eqnarray}
\frac{\Delta E}{E} = \frac{0.25}{\sqrt{E({\rm GeV})}} \oplus 0.01,
\end{eqnarray}
for charged leptons with individual terms added in quadrature.

%------------------------------------------------
% FIG. 3
%------------------------------------------------

\begin{figure}[t!]
\centering\leavevmode
\epsfxsize=4.0in
\epsffile{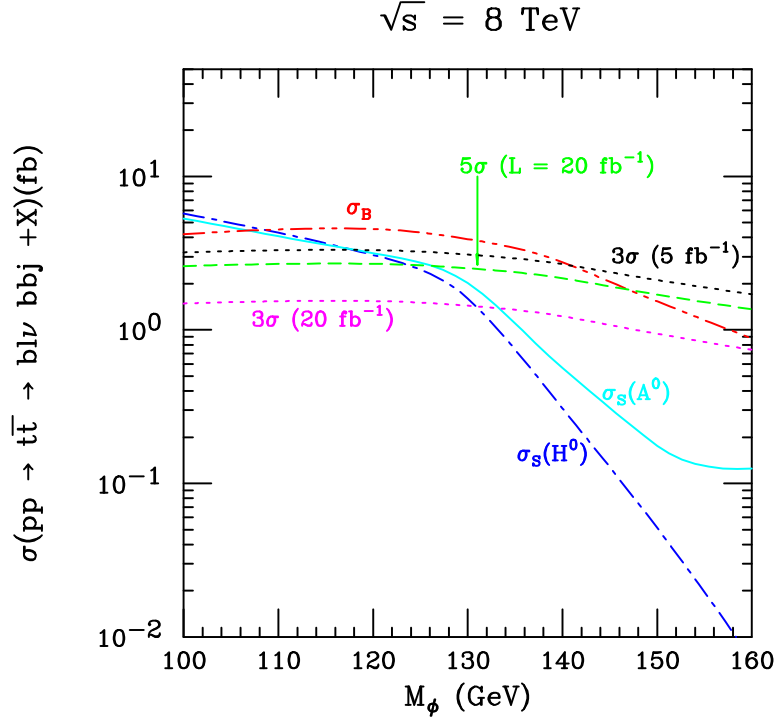}
\caption[]{
The cross section of
$pp \to t\bar{t} \to t c\phi^0 \to b\ell\nu c b\bar{b} +X$
at the LHC with $\sqrt{s} = 8$ TeV, as a function of $M_\phi$.
We present results for Higgs scalar $H^0$ (dash-dot blue) and Higgs
pseudoscalar $A^0$ (solid cyan).
Also shown are the background cross section ($\sigma_B$) 
(dash-dot-dot red) and 
the thresholds for $5 \sigma$ discovery with 20 fb$^{-1}$ (dash green)
of integrated luminosity, or $3 \sigma$
evidence with either 20 fb$^{-1}$ (dot magenta) or 5 fb$^{-1}$ (dot black).
We take the ansatz of Eq.~(5) to set the FCNH coupling strength and
have applied K factors, acceptance cuts, and efficiencies of $b$-tagging
and mistagging.
\label{fig:sigma8} }
\end{figure}

%------------------------------------------------------------------------------
% IV. Discovery Potential at the LHC
%------------------------------------------------------------------------------
\section{Discovery Potential at the LHC}

\begin{figure}[t!]
\centering\leavevmode
\epsfxsize=4.0in
\epsffile{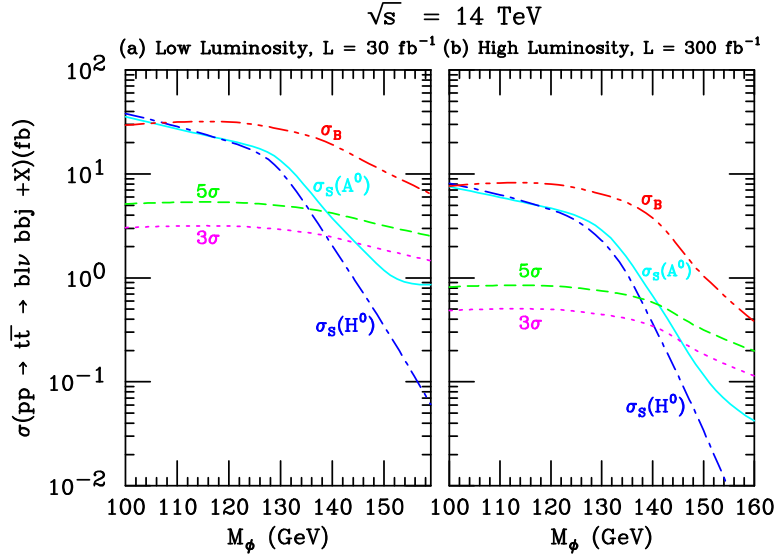}
\caption[]{
The cross section of
$pp \to t\bar{t} \to t c\phi^0 \to b\ell\nu c b\bar{b} +X$
at the LHC with $\sqrt{s} = 14$ TeV, as a function of $M_\phi$.
We present results for Higgs scalar $H^0$ (dash-dot blue) and Higgs
pseudoscalar $A^0$ (solid cyan)
with (a) $L = 30$ fb$^{-1}$ and (b) $L = 300$ fb$^{-1}$.
Also shown are the background cross section(dash-dot-dot red) and
the thresholds for $5 \sigma$ discovery  (dash green) or $3 \sigma$
evidence with  (dot magenta).
We take the ansatz of Eq.~(5) to set the FCNH coupling strength and
have applied K factors, acceptance cuts, and efficiencies of $b$-tagging
and mistagging.
\label{fig:sigma} }
\end{figure}

Our results for the signal and background at the LHC with
$\sqrt{s} = 8$ TeV and $\sqrt{s} = 14$ TeV
are presented in Figs.~\ref{fig:sigma8} and \ref{fig:sigma} respectively.

To estimate the discovery potential at the LHC we include curves that
correspond to the minimal cross section of signal ($\sigma_S$)
required by our discovery criterion described in the following.
We define the signal to be observable
if the lower limit on the signal plus background is larger than
the corresponding upper limit on the background \cite{HGG}
with statistical fluctuations
\begin{eqnarray}
L (\sigma_S+\sigma_B) - N\sqrt{ L(\sigma_S+\sigma_B) } \ge
L \sigma_B +N \sqrt{ L\sigma_B },
\end{eqnarray}
or equivalently,
\begin{equation}
\sigma_S \ge \frac{N}{L}\left[N+2\sqrt{L\sigma_B}\right] \, .
\end{equation}
Here $L$ is the integrated luminosity,
$\sigma_S$ is the cross section of the FCNH signal,
and $\sigma_B$ is the background cross section.
The parameter $N$ specifies the level or probability of discovery.
We take $N = 2.5$, which corresponds to a 5$\sigma$ signal.

For $L\sigma_B \gg 1$, this requirement becomes similar to
\begin{eqnarray}
N_{\rm SS} = \frac{N_S}{\sqrt{N_B}}
 = \frac{L\sigma_S}{\sqrt{L\sigma_B}} \ge 5 \, ,
\end{eqnarray}
where
$N_S$ is the signal number of events,
$N_B$ is the background number of events,
and $N_{\rm SS}$ is the statistical significance, which is
commonly used in the literature.
If the background has fewer than 25 events for a given luminosity,
we employ the Poisson distribution and require that
the Poisson probability for the SM background to fluctuate to this
level is less than $2.85\times 10^{-7}$, i.e. an equivalent probability
to a 5-sigma fluctuation with Gaussian statistics.

%------------------------------------------------
% FIG. 4
%------------------------------------------------

Figure~3 shows the signal and background cross sections for the
CERN Large Hadron Collider with $\sqrt{s} = 8$ TeV.
All tagging efficiencies and K factors discussed above are included.
It is expected that the 2012 run of the LHC will continue at
$\sqrt{s} = 8$ TeV
and accumulate an integrated luminosity of $L \sim 20$ fb$^{-1}$
for each detector at the end of the year \cite{LHC_Lumi}.
As can be seen from the figure, with 20 fb$^{-1}$ of data at 8 TeV
running we can potentially discover this FCNC decay mode for a Higgs with
mass less than $\sim 123$ GeV or we will be able to establish a
$3\sigma$ discrepancy from the Standard Model for masses below $\sim 130$ GeV.
At high masses the
pseudoscalar signal is somewhat larger than the scalar case. This is due to
the absence of $W^+W^-$ and $ZZ$ branching modes in the pseudoscalar decay.

In Fig.~4, we display the signal and background cross sections for the
CERN Large Hadron Collider with $\sqrt{s} = 14$ TeV. We present a lower
luminosity case (LL) with 30 fb$^{-1}$ of data and a long-term high-luminosity
case (HL) with 300 fb$^{-1}$. Cuts and tagging efficiencies are as described
above.

%------------------------------------------------
% FIG. 5
%------------------------------------------------

\begin{figure}[t!]
\centering\leavevmode
\epsfxsize=4.0in
\epsffile{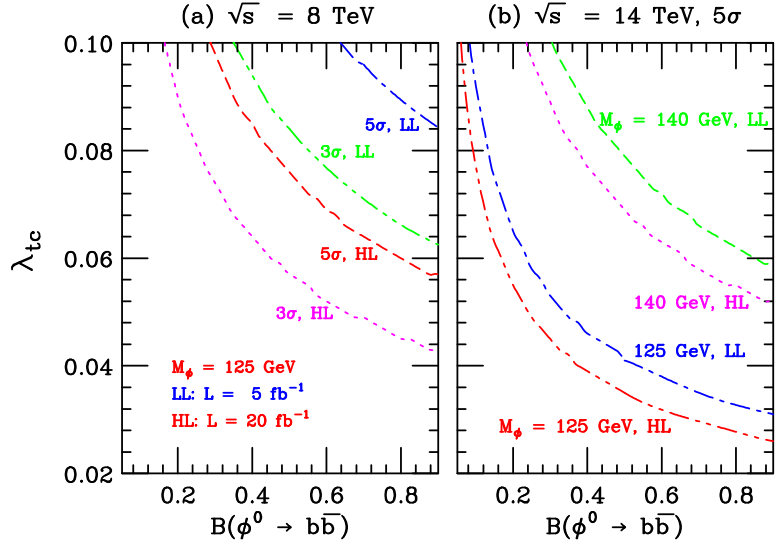}
\caption[]{
The 5$\sigma$ discovery contours at the LHC
in the plane of $[\lambda_{tc}, {\cal B}(\phi^0\to b\bar{b})]$
for
(a) $M_\phi = 125$ GeV
with a low integrated luminosity (LL $= 5$ fb$^{-1}$: dash-dot blue),
and a higher integrated luminosity (HL $= 20$ fb$^{-1}$: dash red)
at $\sqrt{s} = 8$ TeV, and
(b) $M_\phi = 125$ GeV (LL: dash-dot blue, HL: dash-dot-dot red)
or 140 GeV (LL: dash green, HL: dot magenta) with $\sqrt{s} = 14$ TeV and
$L = 30$ fb$^{-1}$ (LL) or $L = 300$ fb$^{-1}$ (HL).
For $\sqrt{s} = 8$ TeV, we also present 3$\sigma$ curves
for $M_\phi = 125$ GeV with LL (dash-dot-dot green) and HL (dot magenta)
where $LL = 5$ fb$^{-1}$ and $HL = 20$ fb$^{-1}$.
The discovery region is the part of the parameter space above the contours.
\label{fig:contour} }
\end{figure}
For $\sqrt{s} = 14$ TeV and $L = 30$ fb$^{-1}$, the range for discovery is
extended up to $M_H \sim 135$ GeV for a scalar coupling or
$M_A \sim 138$ GeV for a pseudoscalar.
With a high luminosity $L = 300$ fb$^{-1}$, these ranges can be
slightly improved to $M_H \sim 138$ GeV and $M_A \sim 142$ GeV respectively.
High luminosity does not extend our range much, owing to (a) the kinematic
limitations as the Higgs mass is increased towards the top mass,
and (b) the higher $p_T$ cuts and lower tagging efficiency we assume.

Figure~5 shows some discovery contours at the LHC for
$M_\phi = 125$ GeV at $\sqrt{s} = 8$ TeV, and
(b) $M_\phi = 125$ GeV or $M_\phi =$ 140 GeV at $\sqrt{s} = 14$ TeV,
as a function of the effective FCNH coupling $\lambda_{tc}$ and
the Higgs to $b\overline{b}$ branching fraction.
These values of Higgs mass are chosen to demonstrate the possible mass range
that might lead to promising FCNH signals.

%-----------------------------------------------------------------------
% V. Conclusions
%-----------------------------------------------------------------------
\section{Conclusions}

It is a generic possibility of particle theories beyond
the standard model to have contributions to tree-level FCNHs,
especially for the third generation quarks.
These contributions arise naturally in models with additional Higgs
doublets, such as the T2HDM, wherein the top quark uniquely couples
to one doublet, or a general 2HDM that has suppressed FCNH for lower generations. 
For a Higgs boson with a mass below the top mass --- a case that seems realized ---
this could engender the rare decay $t \to c \phi^0$.

We investigated the prospects for discovering such a decay at the LHC,
focusing on the channel where $t\overline{t}$ are pair produced and
subsequently decay, one leptonically and one through the FCNH mode.
The primary background for this signal is a $t\overline{t}$ pair
with one standard hadronic decay and the other leptonic.
This background involves one jet mis-tagged as a $b$ jet, which is
much more likely to occur for a $c$ quark than for lighter jets.
Nonetheless, by taking advantage of the available kinematic
information, one can reconstruct the resonances of the signal and
reject much of the background.

Based on a simple geometric ansatz for the size of the FCNC coupling,
namely $\lambda_{tc}/\lambda_t \sim 0.09$, we find
that such a decay mode may be discovered at the LHC for Higgs masses
up to 123 GeV with the current running energy, and up to almost 140 GeV
with the design energy of 14 TeV.
With the emerging scalar object at 125 GeV, this looks rather promising.
We therefore also presented results where the Higgs branching
fraction and FCNC coupling are allowed to vary for fixed masses.

With a new scalar particle very similar to the standard Higgs boson
discovered~\cite{
CMS_Seminar,ATLAS_Seminar,ATLAS_Higgs12,CMS_Higgs12} recently
at the LHC by the ATLAS and CMS experiments,
we look forward to being guided by more new experimental
results as we explore interesting physics of EWSB and FCNH.
While the properties (for example, scalar vs pseudoscalar)
of the Higgs-like object goes under further scrutiny as data accumulates,
perhaps a dedicated FCNH $t\to c\phi^0$ search should be undertaken.

%------------------------------------------------------------------------------
% Acknowledgments
%------------------------------------------------------------------------------
\section*{Acknowledgments}

We are grateful to Paoti Chang for beneficial discussions.
C.K. thanks the Institute of Physics at the Academia Sinica for
hospitality and support during a sabbatical visit.
This research was supported in part
by the U.S. Department of Energy
under Grant No.~DE-FG02-04ER41305,
and by the National Science Council under Grant of Taiwan, R.O.C
under Grant No.~NSC-100-2112-M-001-009-MY3 (Academia Sinica).
WSH thanks the National Science Council for Academic Summit
grant NSC~100-2745-M-002~-002~-ASP.

%------------------------------------------------------------------------------
% Appendix
%------------------------------------------------------------------------------
\section*{Appendix: Comparison of Production Rates}

In this appendix, we present production rates for the FCNH signal and the
dominant background with the same parton distribution functions
(MRST98 Set A~\cite{MRST98}),
the same cuts, and the same efficiencies ($\eps_b = 0.5$,
$\eps_c = \eps_j = 0.01$) and the same K-factors,
used in Ref.~\cite{Aguilar-Saavedra:2000aj}.
For the numerical analysis summarized below, we have adopted the same parameter $g_{tc} = 0.2$ or
$\lambda_{tc} \simeq 0.046$ and the same branching fraction
${\cal B}(H \to b\bar{b}) = 0.7$.
Our results with $H_T(jets+\ell)$ [Table I] are consistently lower,
especially the Higgs signal,
although we have tried to exactly reproduce their method.

Our results become much lower if we take $H_T(jets)$ [Table II] 
to be the scalar sum of $p_T$ for jets only.

\begin{table}[h]
\label{Comparison1}
\caption[]{Comparison of our number of events with results of
Aguilar-Saavedra \& Branco (in parentheses)
calculated with the same cuts, efficiencies, PDFs and scales.
c.f. Table 2 in Ref.~\cite{Aguilar-Saavedra:2000aj}.
In this table, we have chosen $H_T(jets+\ell)$ with jets and leptons.}
\begin{tabular}{|c|c|c|c|c|}
\hline
&\multicolumn{2}{|c|} {Low Luminosity (10 fb$^{-1}$)}
& \multicolumn{2}{|c|} {High Luminosity (100 fb$^{-1}$)}  \\
\cline{2-5}
& Before Cuts & Standard Cuts & Before Cuts & Standard Cuts \\
\hline
Signal & 200 (267) & 46.7 (98.2) & 1630 (2150) & 394 (797)\\
\hline
$t\overline{t}$ & 5491 (7186)& 20.2 (33.2) & 44540 (58230) & 174 (270)\\
\hline
$Wbbjj$ & 58 (77) & 0.232 (0.3) & 476 (644) & 2.00 (2.2)\\
\hline
\end{tabular}

\end{table}

\begin{table}[h]
\label{Comparison2}
\caption[]{Comparison of our number of events with results of
Aguilar-Saavedra \& Branco (in parentheses)
calculated with the same cuts, efficiencies, PDFs and scales.
c.f. Table 2 in Ref.~\cite{Aguilar-Saavedra:2000aj}.
In this table, we have chosen $H_T(jets)$ with only jets.}
\begin{tabular}{|c|c|c|c|c|}
\hline
&\multicolumn{2}{|c|} {Low Luminosity (10 fb$^{-1}$)}
& \multicolumn{2}{|c|} {High Luminosity (100 fb$^{-1}$)}  \\
\cline{2-5}
& Before Cuts & Standard Cuts & Before Cuts & Standard Cuts \\
\hline
Signal & 200 (267) & 30.4 (98.2) & 1630 (2150) & 251 (797)\\
\hline
$t\overline{t}$ & 5491 (7186)& 10.1 (33.2) & 44540 (58230) & 83.9 (270)\\
\hline
$Wbbjj$ & 58 (77) & 0.085 (0.3) & 476 (644) & 0.680 (2.2)\\
\hline
\end{tabular}

\end{table}

%------------------------------------------------
% NEW PAGE
%------------------------------------------------
\newpage

%-----------------------------------------------------------------------
% Bibliography
%-----------------------------------------------------------------------

%-----------------------------------------------------------------------
% END DOCUMENT
%-----------------------------------------------------------------------
\end{document}